\documentclass[submission,copyright,creativecommons]{eptcs}
\usepackage{multicol}
\usepackage{graphics}
\usepackage{graphicx}
\usepackage{subcaption}
\usepackage{enumerate}
\usepackage{array}
\usepackage{booktabs}
\usepackage{url}
\usepackage{xcolor}
 \usepackage{amssymb}

\providecommand{\event}{GCM 2023 Post-Proceedings} 

\usepackage{iftex}

\ifpdf
  \usepackage{underscore}         
  \usepackage[T1]{fontenc}        
\else
 \usepackage{breakurl}           
\fi

\title{Pedagogy of Teaching Pointers in the C Programming Language using Graph Transformations}
\author{Adwoa Donyina
\institute{Computer Science Department\\ Tagliatela College of Engineering\\University of New Haven\\ Connecticut,  USA}
\email{adonyina@newhaven.edu}
\and
Reiko Heckel
\institute{School of Computing and Mathematical Sciences\\University of Leicester\\Leicester, UK}
\email{rh122@leicester.ac.uk}
}
\def\titlerunning{Pedagogy of Teaching Pointers in the C Programming Language using Graph Transformation}
\def\authorrunning{Adwoa Donyina \& Reiko Heckel}
\begin{document}
\maketitle

\begin{abstract}
Visual learners think in pictures rather than words and learn best when they utilize representations based on graphs, tables, charts, maps, colors and diagrams.
%
 We propose a new pedagogy for teaching pointers in the C programming language using graph transformation systems to visually simulate pointer manipulation. In an Introduction to C course, the topic of pointers is often the most difficult one for students to understand; therefore, we experiment with graph-based representations of dynamic pointer structures to reinforce the learning. Groove, a graph transformation tool, is used to illustrate the behaviour of pointers through modelling and simulation. A study is presented to evaluate the effectiveness of the approach. This paper will also provide a  comparison to other teaching methods in this area.
\end{abstract}

\section{Introduction}


Every student has a distinct learning style, which should be enabled by their learning environment to allow them to succeed. In~\cite{learning} the authors evaluate different learning styles in a move towards more student-centered classes. There are visual and non-visual learners; however visual learners tend to gain higher comprehension~\cite{visuallearner}. Within the realm of visual teaching techniques there are different depths, from using visual aids such as PowerPoint slides via static more or less formal representations such as pictures, maps, diagrams and charts, to dynamic and interactive visualisations. 

We are proposing a visual learning pedagogy using graph transformations for the modelling and simulation of pointer manipulations in C. This is often the most difficult topic for students to understand in an introductory C programming course. 
We report on an experiment in utilising a graph-based representation of dynamic pointer structures during a revision session at the end of the course and compare the test results of this class to those of an otherwise equivalent class a year earlier.




This paper is organized as follows. In 
Section~\ref{Background} we  discuss background on the traditional methods of teaching C pointers at university and briefly introduce concepts needed to understand our proposed pedagogy. In Section~\ref{pointer} we present our approach of representing pointer manipulations using graph transformation. In Section~\ref{Simulation} we illustrate it by a simulation based on an example. In Section~\ref{Experiment} we present results from an experiment using the approach during the revision stage of a programming course. In Section~\ref{Related} we compare and contrast related work. In Section~\ref{Conclusion} we will conclude the paper and discuss future work. 

\section{Background}\label{Background}
In many universities, C  is the first programming language on the curriculum. The Computer Science and Cyber Security programme at the University of New Haven (UNH) teaches computer languages in the order of C, C++, Python, Java. In more advanced classes students utilize these languages to study data structures and algorithms, and more specific subjects. The initial programming course teaches many fundamental skills to help students starting their programming career. Learning to program is difficult, and pointer manipulation poses particular challenges to students~\cite{industralExample}.
Therefore, there is a need to explore new pedagogical approaches to support teaching the hardest concepts in C programming.

The Introduction to C Programming class at UNH is based on Chapters 1-13 of the textbook~\cite{afischer}. Pointers are introduced at the end of the term, to ensure that students understand the basic topics prior to being introduced to this machine-oriented way of thinking.
\begin{multicols}{2}\label{syllabus}
\begin{enumerate}
  \item Computers and Systems
  \item Programs and Programming 
  \item Fundamental Concepts
  \item Objects, Types and Expressions
  \item Using Functions and Libraries
  \item More Repetition and Decision
  \item Using Numeric Types
  \item Trouble with Numbers
  \item Program Design
  \item An Introduction to Arrays
  \item Character Data and Enumerations
  \item An Introduction to Pointers
\item Strings  
\end{enumerate}
\end{multicols}
 For many years pointer manipulation was taught only using the textbook~\cite{afischer} as a resource. This uses informal illustrations of pointer structures and their changes during program execution but naturally they are restricted to a few examples and do not allow to simulate alternatives based on different input states or to show the effects of erroneous code. To improve the teaching of pointer manipulation by visual means  we propose to use graph transformation~\cite{hartmut1,HT2020}, in particular  Groove~\cite{groove}, as a pedagogical aid.

Graph transformation is a \emph{rule-based} approach which represents procedural knowledge in terms of a set of ``IF-THEN" rules. These rules define the preconditions and effects of the basic activities. This form of modelling is in contrast to the  \textit{control-oriented} approach, which focuses on the ordering of events.  

Formally,  graph transformation rules consist of left hand sides, right hand sides and negative application conditions (NACs), all of which are different graphs; however Groove combines them in one graph with colour coding~\cite{groove}. The tool also supports advanced concepts such as graph constraints, negative application conditions, multi objects and patterns, and control flow specifications as well as the analysis by model checking.

\section{Pointer Manipulation using Graph Transformation}\label{pointer}
In addition to the traditional means of teaching pointers in the C programming language, we are proposing a new pedagogy using graph transformation systems to visually simulate pointer manipulation. The Groove~\cite{groove} graph transformation tool is used to create and simulate executable pointer models.
 
\begin{figure}[h!]
  \begin{center}
  \includegraphics[width=12cm]{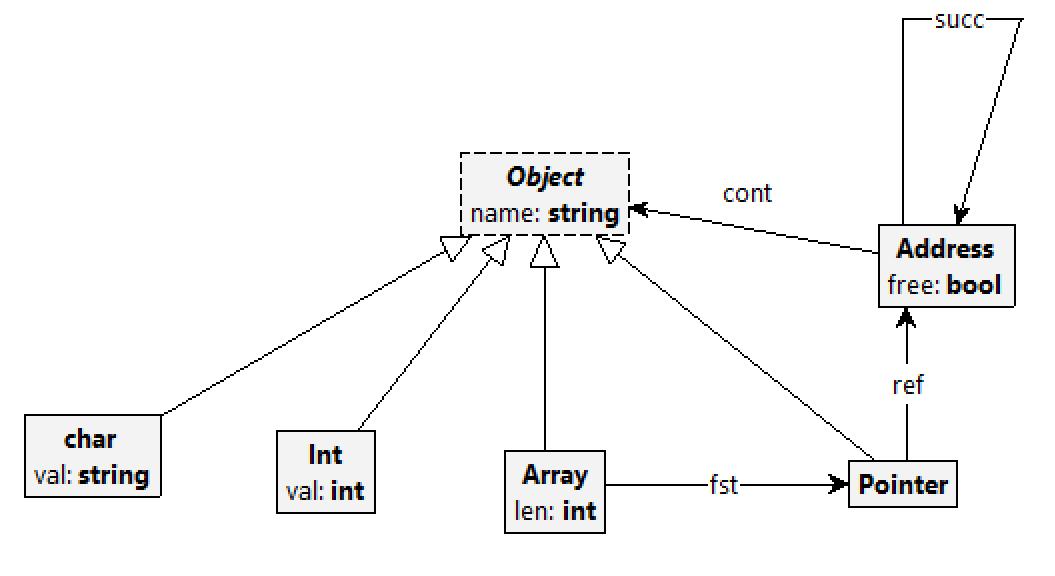}
    \end{center}
    \vspace{-20pt}
  \caption{Type graph for pointer structures}
  \label{typegraph}
\end{figure}

As part of a our experiment, we designed a type graph to model basic C pointer structures as shown in Figure~\ref{typegraph}. This type graph represents how C pointers relate to addresses and objects, such as array, int or char, that may reside at those addresses. Addresses represent memory cells linked by \emph{succ} edges if they are consecutive.
Each pointer may refer to an address and each address may contain an object. An array is a consecutive sequence of addresses containing objects, and the Array node has a \emph{fst} edge to the pointer referencing the first address of the array. The array's length is specified using attribute \emph{len}. We use the Object type's name attribute to refer to a variable in the program code, but this is information is never used in the rules describing the operations and serves only to improve readability of the generated graphs. In particular, neither addresses nor pointers directly represent variables. Our type graph does not account for more advanced concepts such as objects of different size or the distinction between heap and stack memory, which were not covered in our introductory module.
 
An important design decision is the explicit modelling of addresses representing memory cells and their consecutive organisation in memory (by means of \emph{succ} edges). Their status as free or allocated  is recorded by the Boolean attribute \emph{free} to reflect the effect of C memory functions \texttt{malloc()} and \texttt{free()}. 
This is a lower level of representation than the obvious model of pointer structures where pointers are edges between objects; however, this implementation-oriented model also allows us to capture operations such as dereferencing \texttt{*} and address-of \texttt{\&}. As importantly, we can show situations where the structure violates referential integrity, e.g., when a pointer refers to an address which is either free or holds no object, potentially leading to inconsistencies we would like to illustrate in the execution of an erroneous program.
We define graph constraints over the type graph to represent both referential integrity and basic constraints on the model. For the latter:
\begin{itemize}
    \item \emph{fst} edges have *-to-1 cardinality; 
    \item \emph{ref} edges have *-to-0..1 cardinality, where a pointer with no \emph{ref} edge has value \emph{null};
    \item \emph{cont} edges have 0..1-to-0..1 cardinality, i.e., while each object in memory has a unique address we allow partial representations where the address of, for example, a pointer object, is not relevant, and an address can only be the first address of at most one object;
    \item \emph{succ} edges have 0..1-to-0..1 cardinality, i.e., each address apart from the first / last has a predecessor / successor; specifically, \emph{succ} edges should form a single chain representing a linear address space.
\end{itemize}

\begin{figure}[h!]
\centering
\begin{subfigure}{0.4\textwidth}
\centering
\includegraphics[width = 5cm]{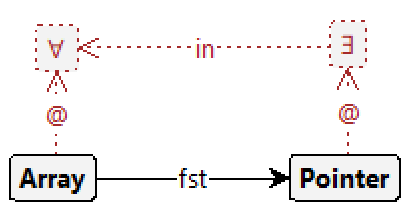}
\caption{isWFfstEx, using nested universal and existential quantifiers
}
\label{isWFfstEx}
\end{subfigure}
\quad
\begin{subfigure}{0.4\textwidth}
\centering
\includegraphics[width = 5.5cm]{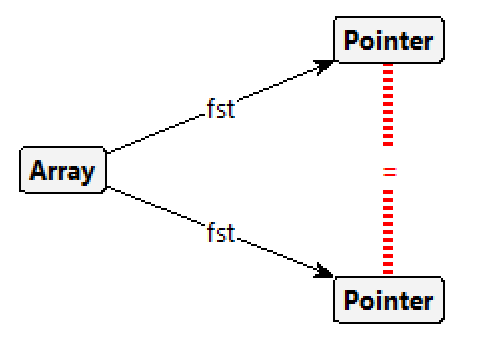}
\caption{notWFfstToV, using red inequality constraint edge labelled = 
}
  \label{notWFfstToV}
\end{subfigure}
\begin{subfigure}{0.45\textwidth}
\centering
\includegraphics[width = 5.5cm]{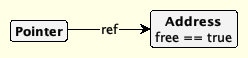}
\caption{notRIrefTofree 
}
\label{notRIrefTofree}
\end{subfigure}
\quad
\begin{subfigure}{0.45\textwidth}
\centering
\includegraphics[width = 7.5 cm]{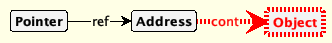}
\caption{notRIrefWOcont 
}
  \label{notRIrefWOcont}
\end{subfigure}

\caption{Graph constraints for well-formedness (WF) and referential integrity (RI)}
\label{gc}
\end{figure}

These constraints must always be satisfied for an instance graph to be \emph{well-formed}, i.e., model a pointer structure. They can be formally specified using graph constraints such as those in Figure~\ref{isWFfstEx} (for all Array nodes there exists an outgoing \emph{fst} edge to a Pointer node) and Figure~\ref{notWFfstToV} (there is an Array node with two outgoing \emph{fst} edges to different Pointer nodes). Groove~\cite{groove} supports quantified graph elements using forall ($\forall$) and exist ($\exists$) nodes. Every node referring to these by an \emph{@} edge is correspondingly quantified, while edges are assigned implicitly to the quantifier of their source or target, whichever is higher, with nesting of quantifiers shown by \emph{in} edges. Hence, the constraint in Figure~\ref{isWFfstEx} expresses the first-order formula \emph{$\forall$ a:Array $\exists$ p:Pointer, f:fst. src(f) = a $\wedge$ tar(f) = p}.
The constraint in Figure~\ref{notWFfstToV} uses a pattern with a negated equality edge indicating that the two Pointer nodes must be matched to different nodes in the graph.
Then, the formula \emph{G (isWFfstEx \& ! notWFfstToV)} defines an invariant for the cardinality of \emph{fst} edges as stated in the first condition above where $G$ is the temporal logic operator stating that the formula holds in all states.
Analogously, we defined the other three cardinalities. 

In order to represent a structure that is \emph{correct}, i.e., respects referential integrity, we require that a \emph{ref} edge never points to an address that is
    free (not allocated) nor 
    has no \emph{cont} edge (does not contain data).

Graph constraints for these two conditions are shown in Figures~\ref{notRIrefTofree} and \ref{notRIrefWOcont} respectively, jointly invoked by the formula \emph{G (! notRIrefTofree \& ! notRIrefWOcont)}. The second constraint uses a negative condition (in red) to indicate the absence of an object at the address. This captures the case where a pointer refers to an address that is not in use, so accessing data from this location leads to a logical error.

Well-formedness should be invariant while correctness is a property of a graph that may be violated by simulations modelling the execution of erroneous programs, which is pedagogically valuable. Either formula can be validated using Groove's model checker.

\begin{figure}[h!]
\centering
\begin{subfigure}{0.49\textwidth}
\centering
\includegraphics[width = 6cm]{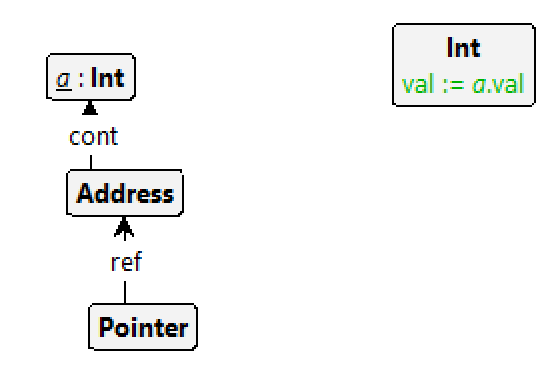}
\caption{Copy value to int object}
\label{copyReferent}
\end{subfigure}
\begin{subfigure}{0.20\textwidth}
\centering
\includegraphics[width = 1.20cm]{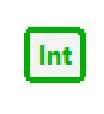}
\caption{New int}
  \label{newInt}
\end{subfigure}
\begin{subfigure}{0.20\textwidth}
\centering
\includegraphics[width = 2cm]{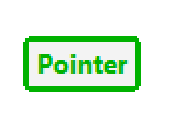}
 \caption{New pointer}
  \label{newPointer}
\end{subfigure}
\caption{Basic rules with int \& pointers}
\label{GTbasic}
\end{figure}

\begin{figure}[h!]
\centering
\begin{subfigure}{0.49\textwidth}
\centering
\includegraphics[width = 6cm]{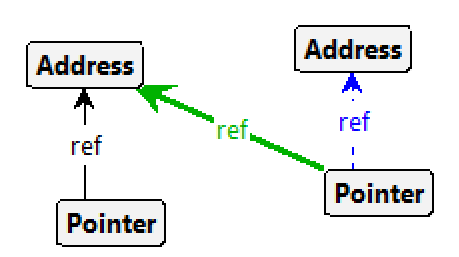}
\caption{Not null pointer assigned address of another}
 \label{pointerReferent}
\end{subfigure}
\begin{subfigure}{0.49\textwidth}
\centering
\includegraphics[width = 5cm]{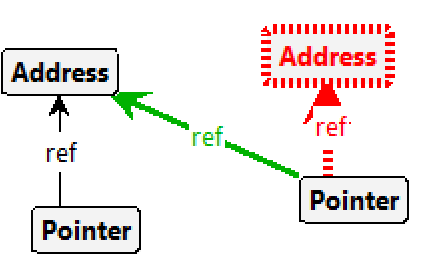}
\caption{Null Pointer assigned address of another}
  \label{nullPointerReferent}
\end{subfigure}
\caption{A pointer is assigned the address of another pointer}
\label{GT-assignPointertoPointer}
\end{figure}

\begin{figure}[h!]
\centering
\begin{subfigure}{0.49\textwidth}
\centering
\includegraphics[width = 5cm]{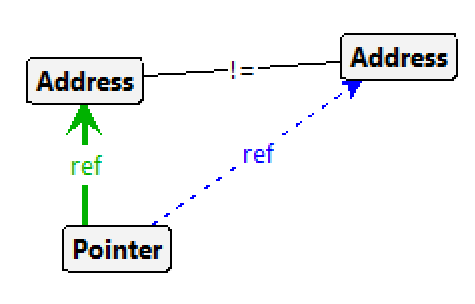}
  \caption{Pointer is assigned new address}
 \label{pointerAssignednewAddress}
\end{subfigure}
\begin{subfigure}{0.49\textwidth}
\centering
\includegraphics[width = 5cm]{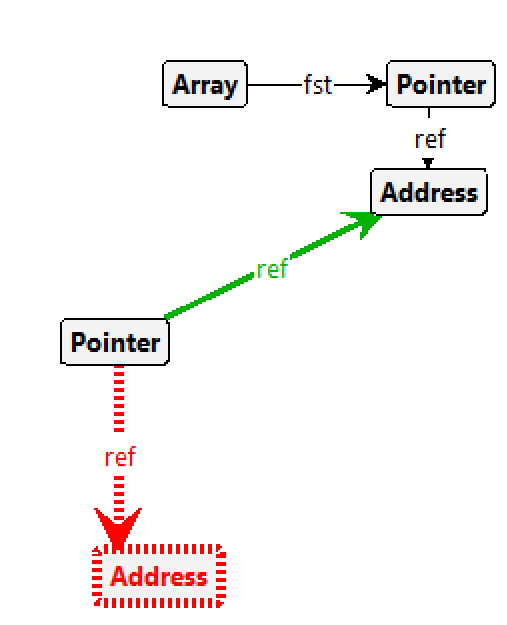}
  \caption{Null Pointer is assigned address of array}
  \label{pointerArray}
\end{subfigure}
\caption{Pointer is assigned address of array}
\label{GT-assignPointertoAddressArray}
\end{figure}

\begin{figure}[h!]
\centering
\begin{subfigure}{0.49\textwidth}
\centering
\includegraphics[width = 8cm]{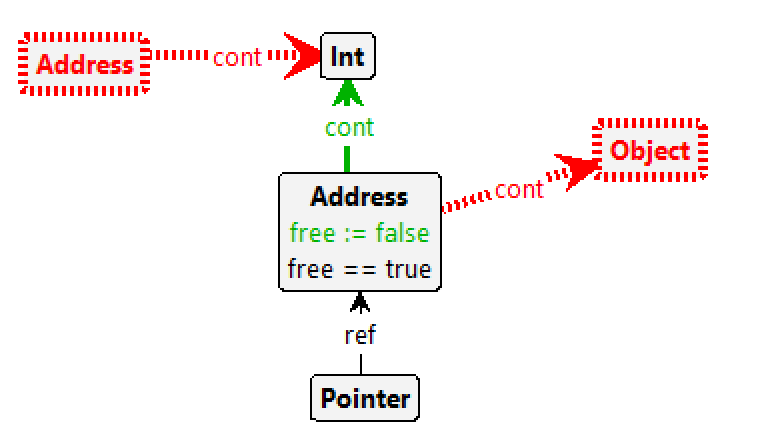}
  \caption{Int is stored at (free) address referred to by pointer}
  \label{pointerInt}
\end{subfigure}
\begin{subfigure}{0.49\textwidth}
\centering
\includegraphics[width = 8cm]{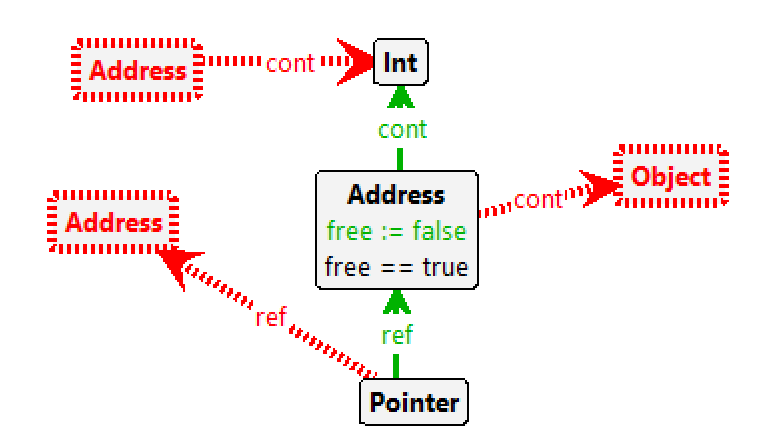}
  \caption{Null pointer is assigned address where int is stored}
 \label{nullpointerInt}
\end{subfigure}
\caption{Int is stored at address and referred to by pointer}
\label{GT-assignPointertoInt}
\end{figure}

 A number of graph transformation rules encode basic operations of pointer manipulations. The rules use the Groove notation where left- and right-hand sides are integrated into one graph, with colours representing which elements are required and preserved (solid black outline), deleted (dashed blue outline), and created (bolder green outline). Forbidden elements (red dotted outline) represent negative application conditions.
\begin{itemize}
\item
The rule in Figure~\ref{copyReferent} copies the value of the \texttt{int} referred to by a pointer into another \texttt{int} object. For instance, if \texttt{int *pt} and \texttt{int s}, the rule realises  \texttt{s = *pt}. 
\item The rule in Figure~\ref{newInt} creates an \texttt{int} object.
\item The rule in Figure~\ref{newPointer} creates a \texttt{null} pointer.
\item The rule in Figure~\ref{pointerReferent} assigns to a pointer the address of another pointer. For instance, if \texttt{int *pt} and \texttt{int *pt2} are both not \texttt{null}, the rule realises \texttt{pt2=pt}.
\item The rule in Figure~\ref{nullPointerReferent} is similar to the one in Figure~\ref{pointerReferent}, however it assigns the address to a \texttt{null} pointer. For instance, if \texttt{int *pt} is not \texttt{null} but \texttt{int *pt2} is \texttt{null}, the rule realises \texttt{pt2=pt}. This is expressed as a negative application condition, shown in red dotted outline, specifying the absence of a \emph{ref} edge to an Address node.
%
\item The rule in Figure~\ref{pointerAssignednewAddress} assigns to a pointer a different address.
\item The rule in Figure~\ref{pointerArray} assigns to a pointer the address of (the first element of) an existing array. For instance, if \texttt{int a[]=\{4,7,9\}} and \texttt{int *pt} is \texttt{null}, the rule realises \texttt{pt=a} and \texttt{pt=a[0]}. Similar rules are defined to access other positions in the array using the \emph{succ} relation.
\item The rule in  Figure~\ref{pointerInt} assigns to a pointer the address of an existing \texttt{int} object. After executing this rule, the pointer refers to the \texttt{int}'s address.
\item The rule in  Figure~\ref{nullpointerInt} assigns to a \texttt{null} pointer an address which now contains an existing int object. For instance, if \texttt{int b=25} and \texttt{int *pt} is \texttt{null}, the rule realises \texttt{*pt=b}.
\end{itemize}
This set of rules is enough to illustrate the scenario in Section~\ref{Simulation} but not complete for the entire range of pointer operations in the C language. 

\begin{figure}[h!]
  \begin{center}
  \includegraphics[width=15cm]{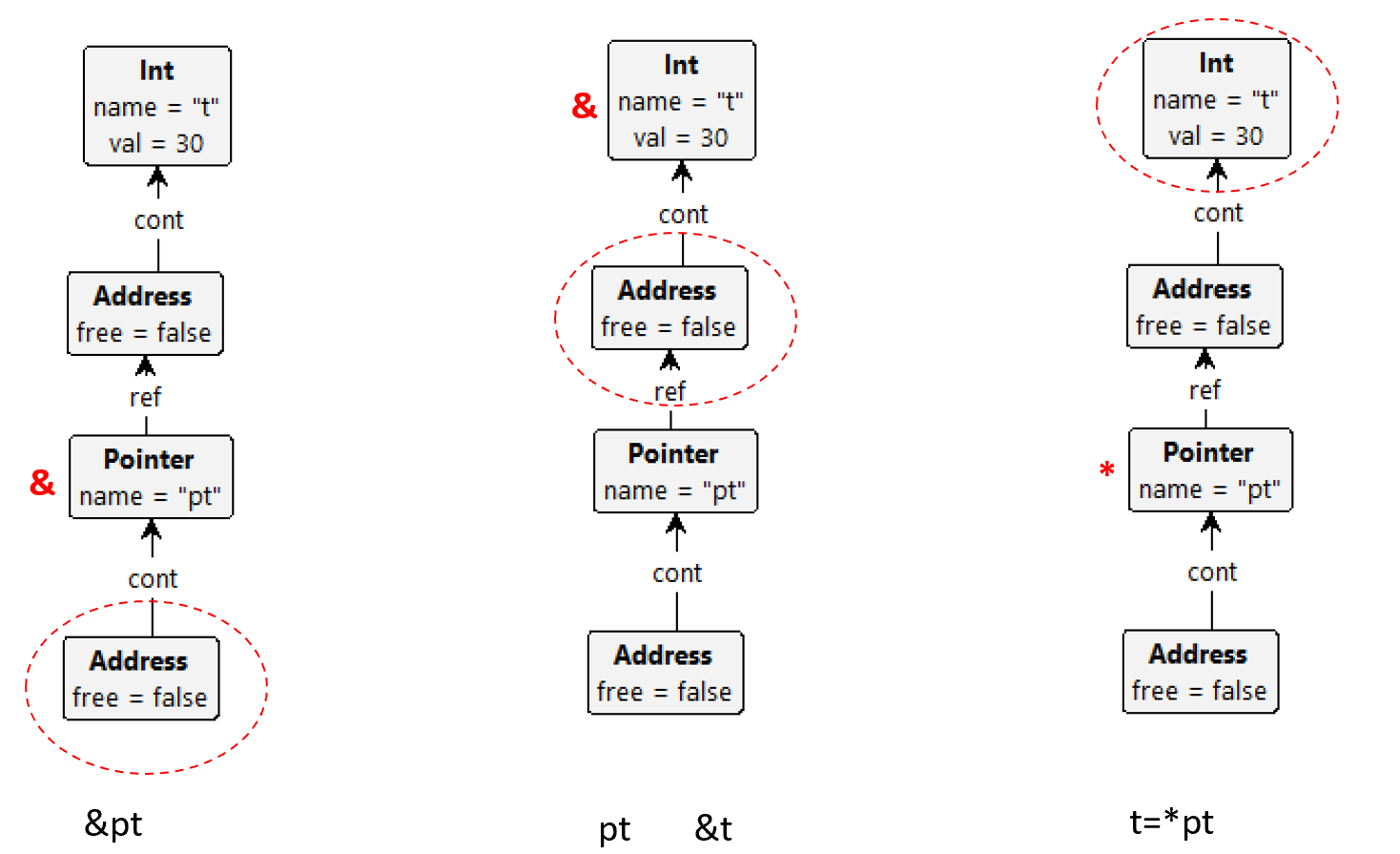}
    \end{center}
  \caption{Effect of \& vs * Notation}
  \label{notation}
\end{figure}

In C code, pointer operations are denoted by the \texttt{\&} and \texttt{*} symbols. The asterisk \texttt{*}, apart from being used for pointer declaration, returns the object at the address referenced by a pointer. In terms of our model this means to navigate from the Pointer to the Object via the Address node along \emph{ref} and \emph{cont}.
Instead, \texttt{\&} returns the address of an object, navigating \emph{cont} in the reverse. 
Hence, if \texttt{int *pt} defines a pointer to the address of  \texttt{int t=30} (or, more compactly: \texttt{int t=30; int *pt=\&t}), then 
\begin{itemize}
    \item \texttt{pt} returns the address referenced by the pointer, or null if the pointer is undefined;
    \item \texttt{\&pt} returns the address of the pointer as an object;
    \item \texttt{*pt} returns the int value 30;
   \item \texttt{pt} and \texttt{\&*pt} both return the address of the int.
\end{itemize}
Hence,  if \texttt{int s} and \texttt{int t} are defined then \texttt{s = *pt} assigns the value of the int referred to by pointer \texttt{pt} to an int variable \texttt{t}. This use of \texttt{\&} and \texttt{*} is illustrated in Figure~\ref{notation}. The dotted circle annotations shows what the notation is referring to, i.e., the pointer's content address and int value, respectively.

\begin{figure}[h!]
  \begin{center}
  \includegraphics[width=13cm]{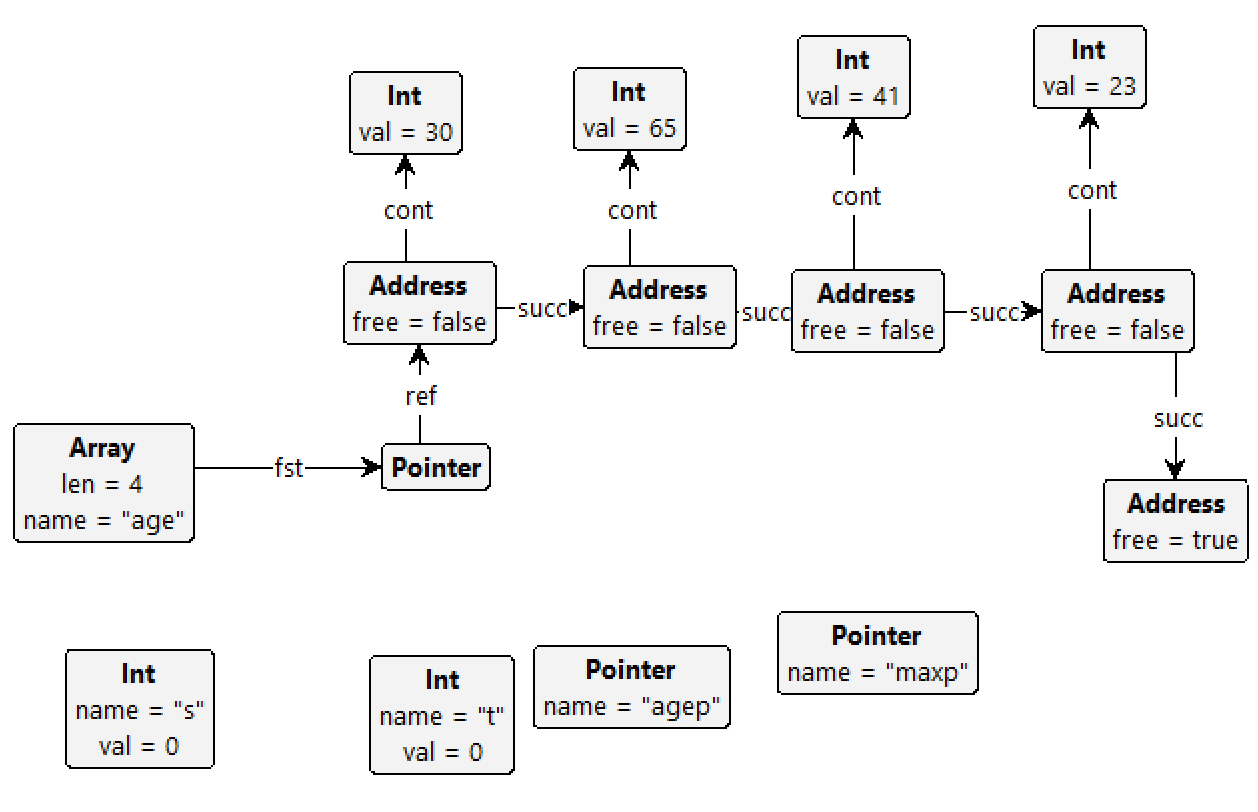}
    \end{center}
    \vspace{-20pt}
  \caption{Instance graph representing start graph of simulation}
  \label{startgraph}
\end{figure}

The result of executing the following C code is represented by the instance graph in Figure~\ref{startgraph}. 
\texttt{\newline
int s = 0; \newline
int t = 0; \newline
int age[] = \{ 30, 65, 41, 23 \}; \newline
int *agep, *maxp; 
}

In the bottom left are objects of type int represented by the variables \texttt{s} and \texttt{t}, currently unconnected. The \texttt{age} array refers to the pointer holding the address of the first element of the array. This and the three subsequent addresses contain the corresponding int values of the initial array. 

\section{Simulation}\label{Simulation}

The instance graph defined in Figure~\ref{startgraph} is the start graph for our simulations, whose runs were utilized to illustrate the following code.
\newline
\texttt{s=*age;}  OR \texttt{ s=age[0];} \newline
\texttt{agep=age;} \newline
\texttt{agep= \&age[3];} \newline
\texttt{*maxp=t;}

\begin{figure}[h!]
  \begin{center}
  \includegraphics[width=13cm]{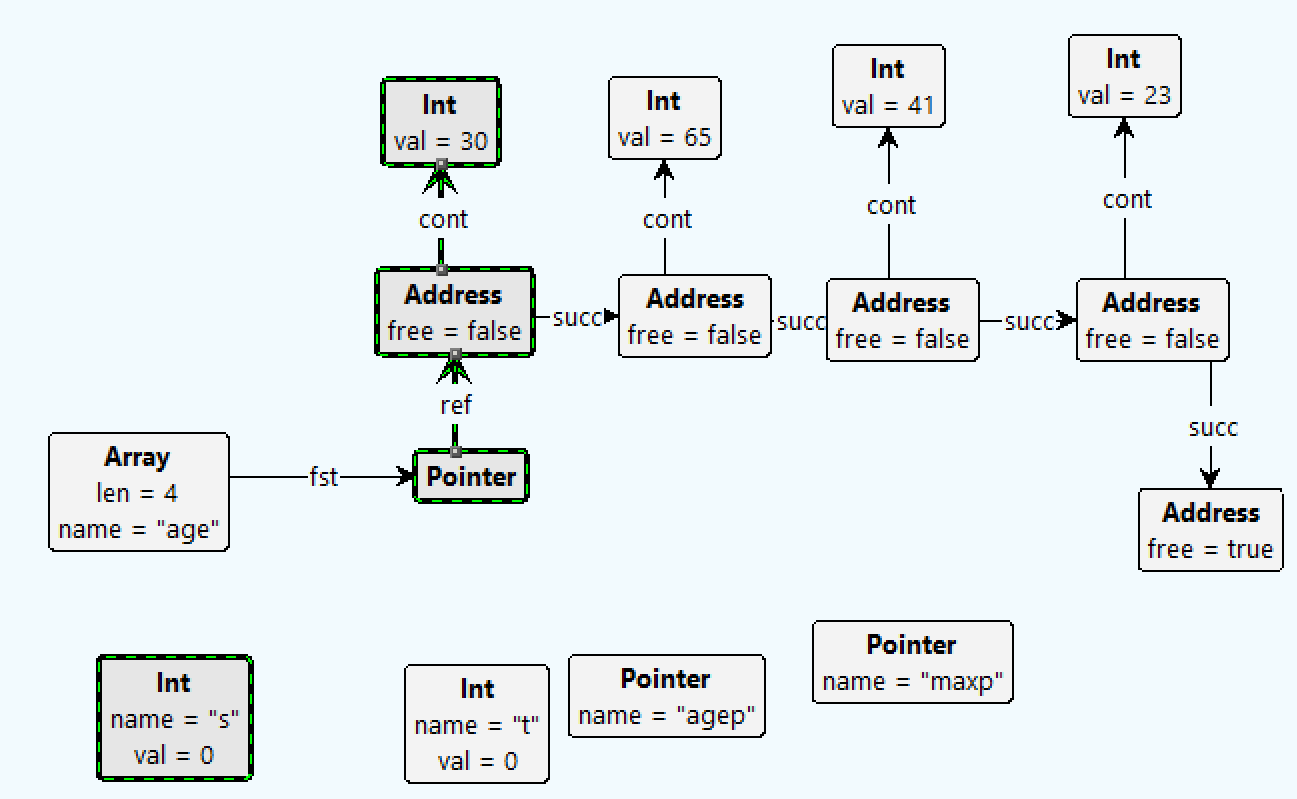}
    \end{center}
    \vspace{-20pt}
  \caption{Match of rule in Fig.~\ref{copyReferent}: copying value into object}
  \label{step1-match}
\end{figure}

\begin{figure}[h!]
  \begin{center}
  \includegraphics[width=13cm]{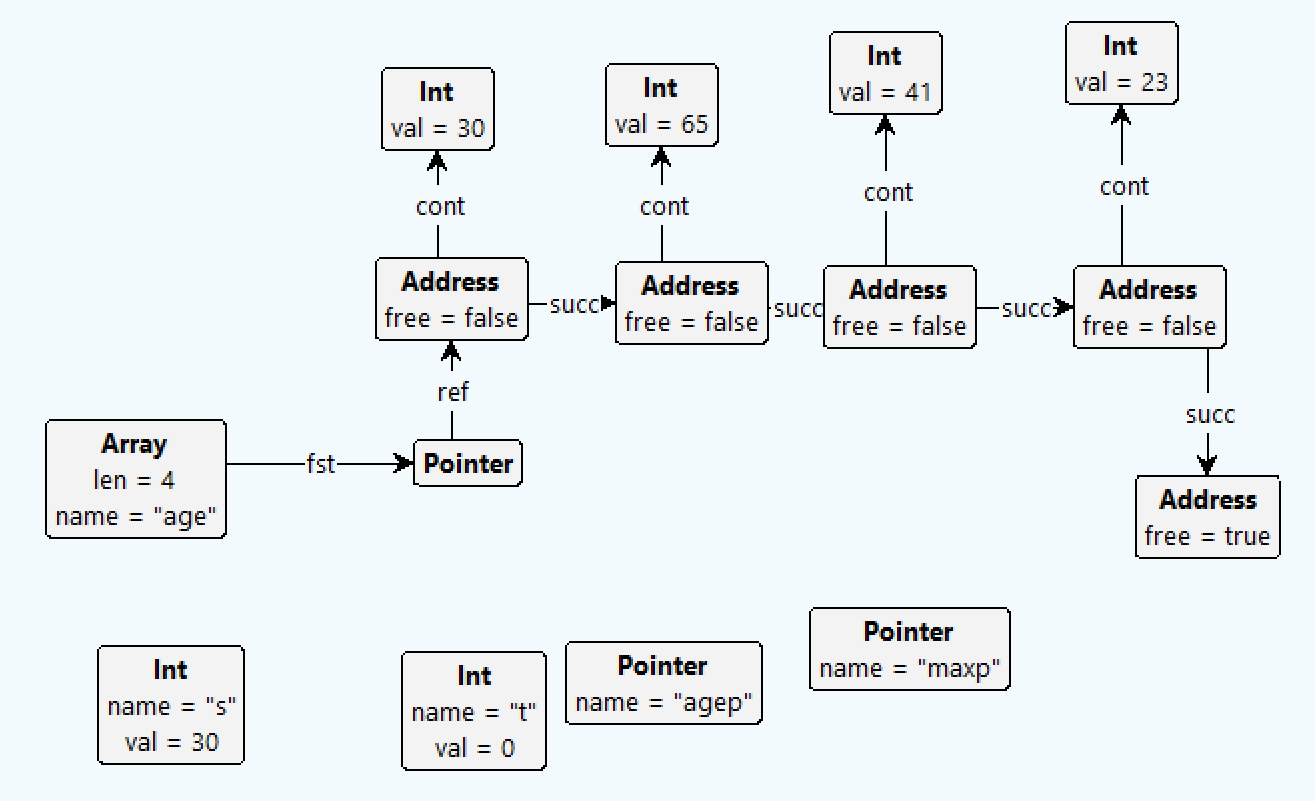}
    \end{center}
    \vspace{-20pt}
  \caption{Rule in Fig.~\ref{copyReferent} applied for \texttt{s=*age;}}
  \label{step1-applied}
\end{figure}

\begin{figure}[h!]
  \begin{center}
  \includegraphics[width=13cm]{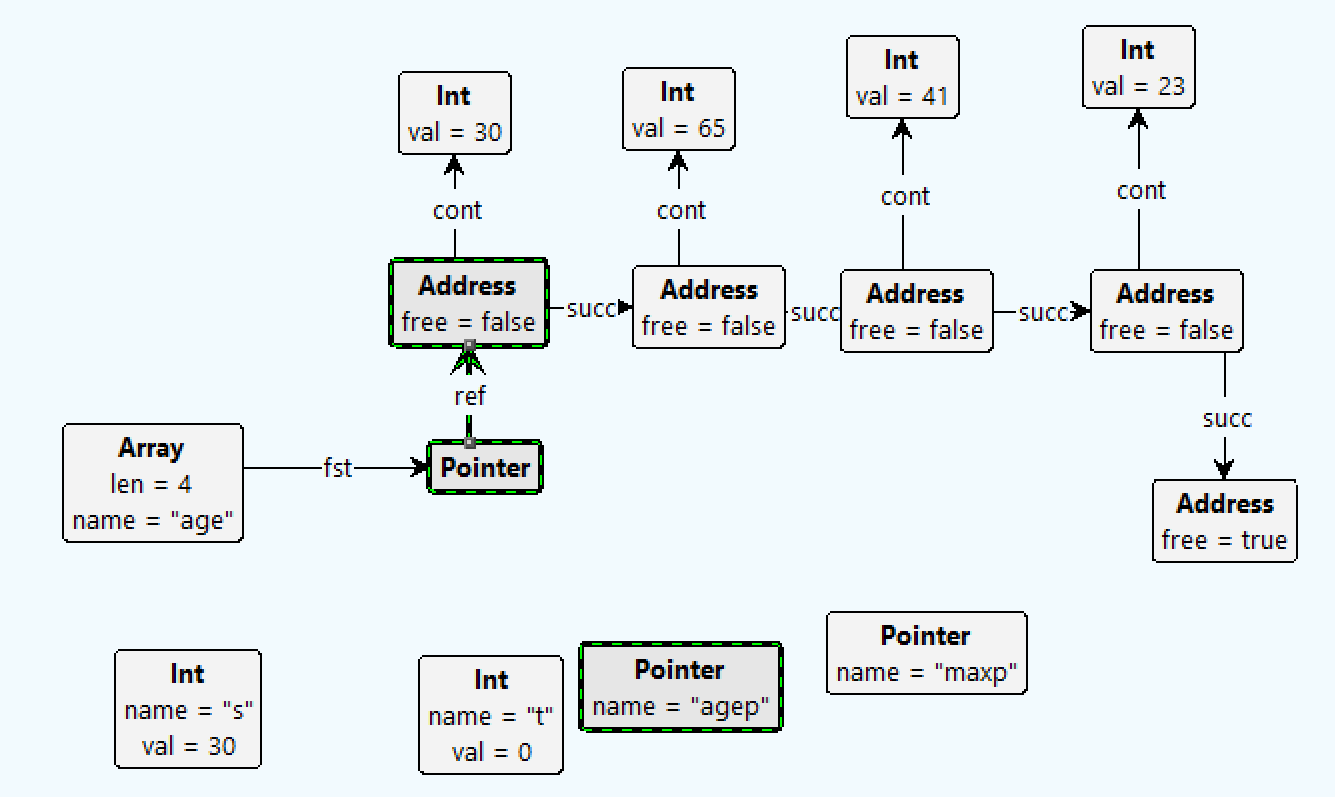}
    \end{center}
    \vspace{-20pt}
  \caption{Match of rule in Fig.~\ref{nullPointerReferent}: assigning a null pointer to another pointer's referent}
  \label{step2-match}
\end{figure}

\begin{figure}[h!]
  \begin{center}
  \includegraphics[width=13cm]{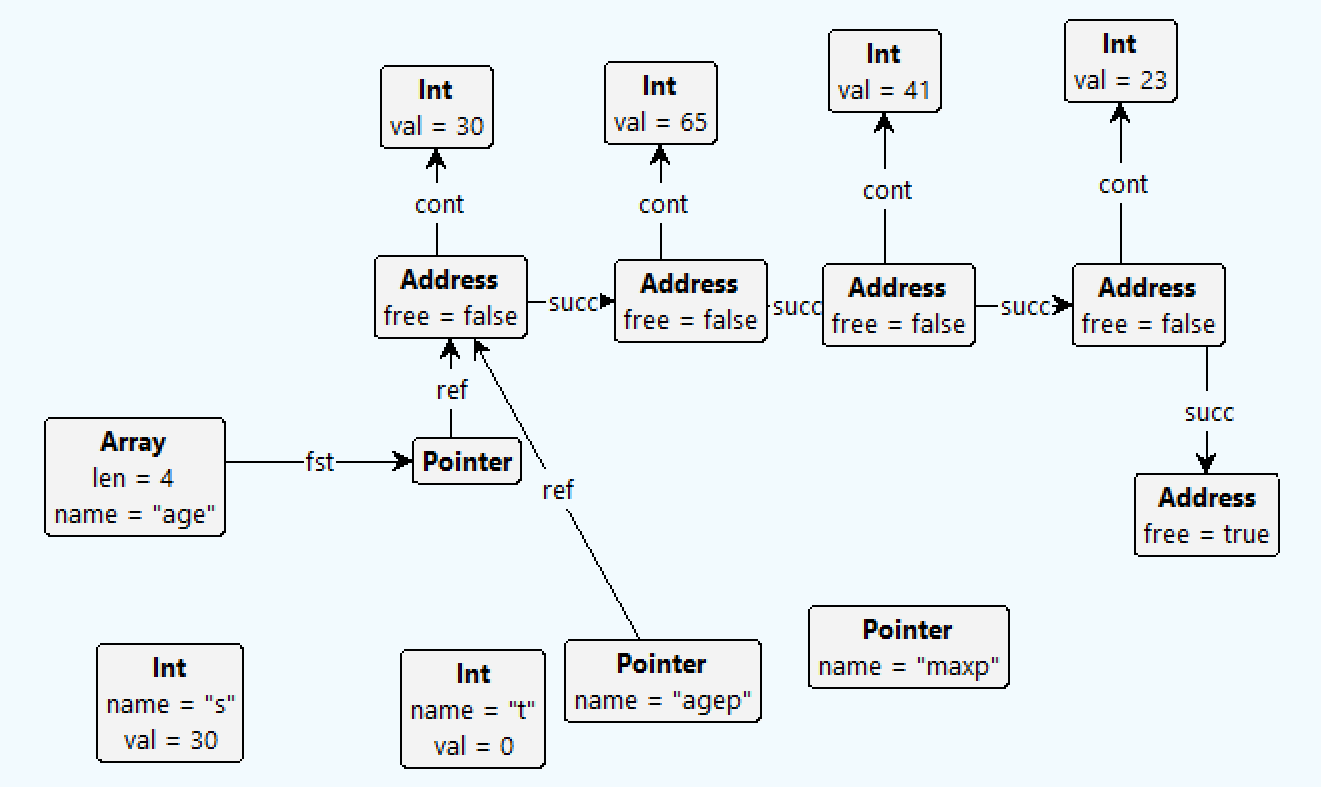}
    \end{center}
    \vspace{-20pt}
  \caption{Rule in Fig.~\ref{nullPointerReferent} applied for \texttt{agep=age;}}
  \label{step2-applied}
\end{figure}

\begin{figure}[h]
  \begin{center}
  \includegraphics[width=13cm]{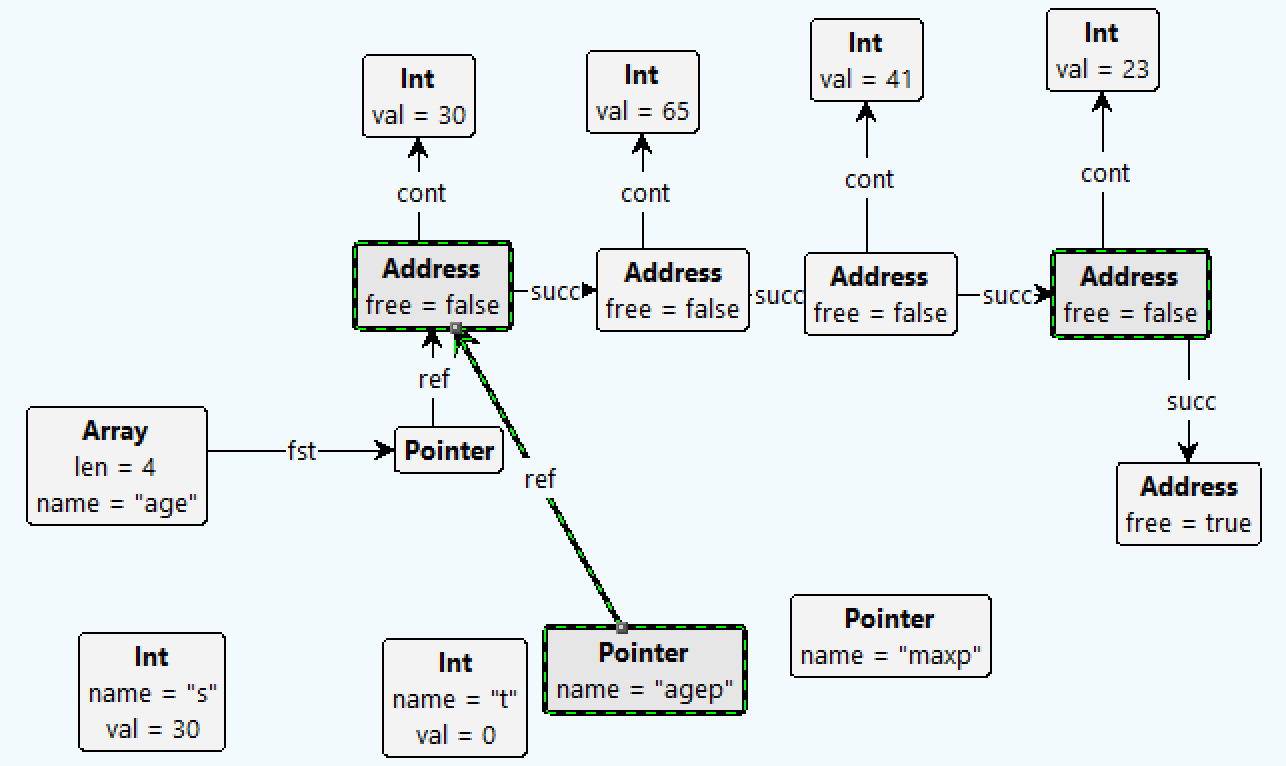}
    \end{center}
    \vspace{-20pt}
  \caption{Match of rule in Fig.~\ref{pointerAssignednewAddress}: assigning a pointer to new address}
  \label{step3-match}
\end{figure}

\begin{figure}[h!]
  \begin{center}
  \includegraphics[width=13cm]{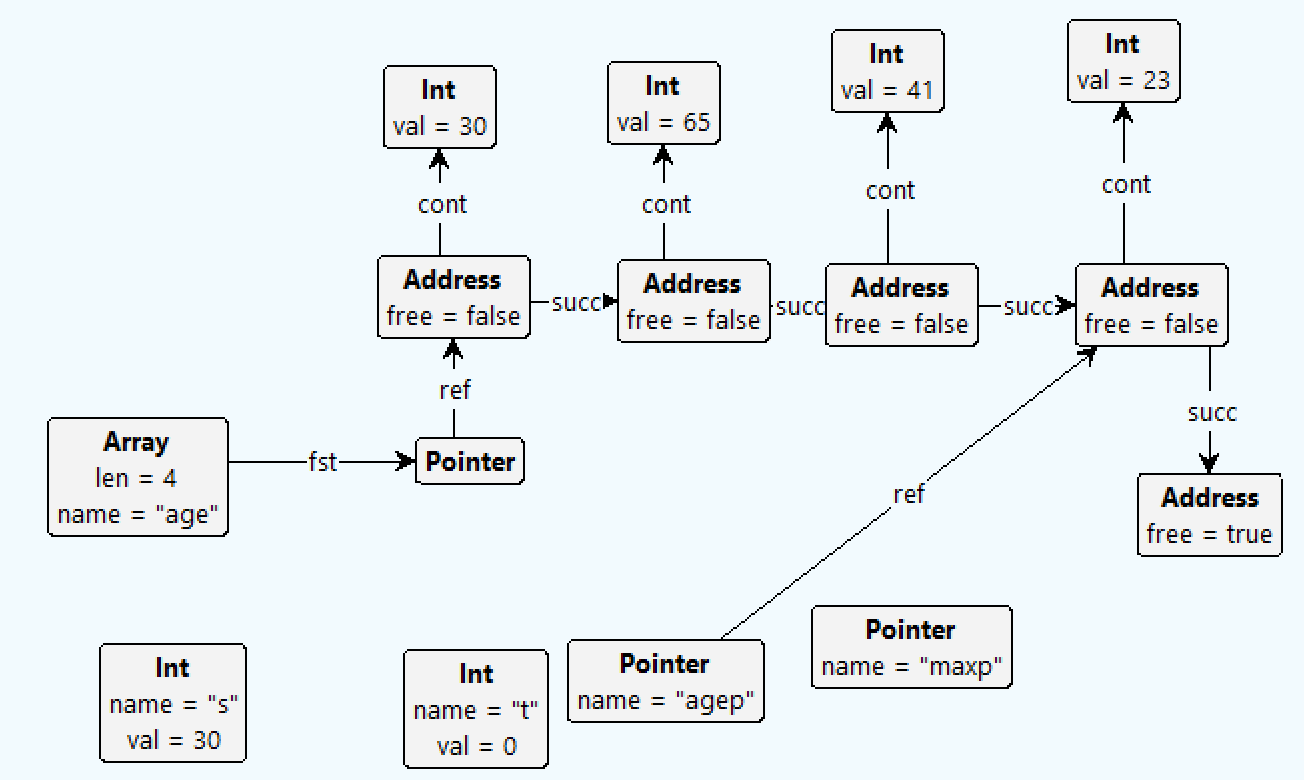}
    \end{center}
    \vspace{-20pt}
  \caption{Rule in Fig.~\ref{pointerAssignednewAddress} applied for \texttt{agep= \&age[3];}}
  \label{step3-applied}
\end{figure}

\begin{figure}[h!]
  \begin{center}
  \includegraphics[width=13cm]{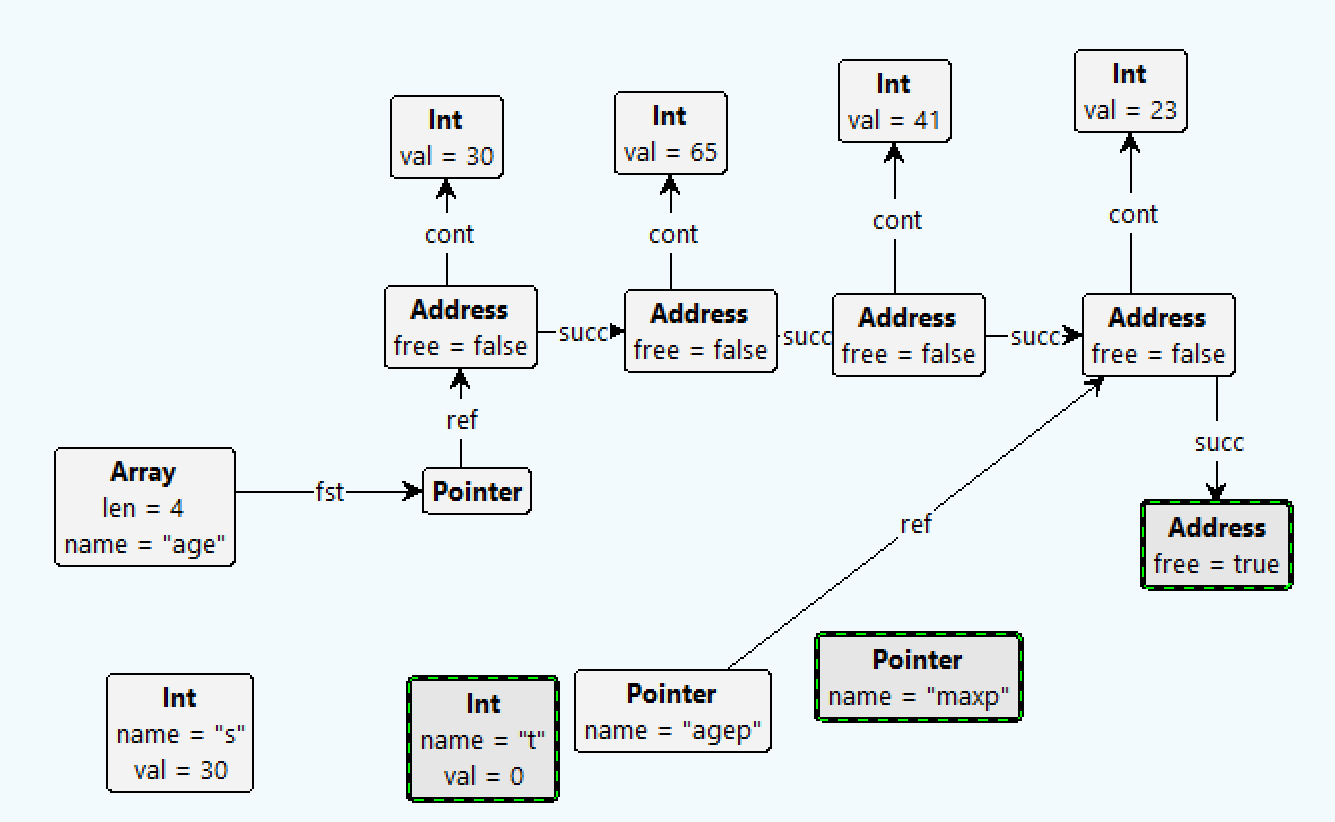}
    \end{center}
    \vspace{-20pt}
  \caption{Match of rule in Fig.~\ref{nullpointerInt}: null pointer is assigned to the address of a int}
  \label{step4-match}
\end{figure}

\begin{figure}[h!]
  \begin{center}
  \includegraphics[width=13cm]{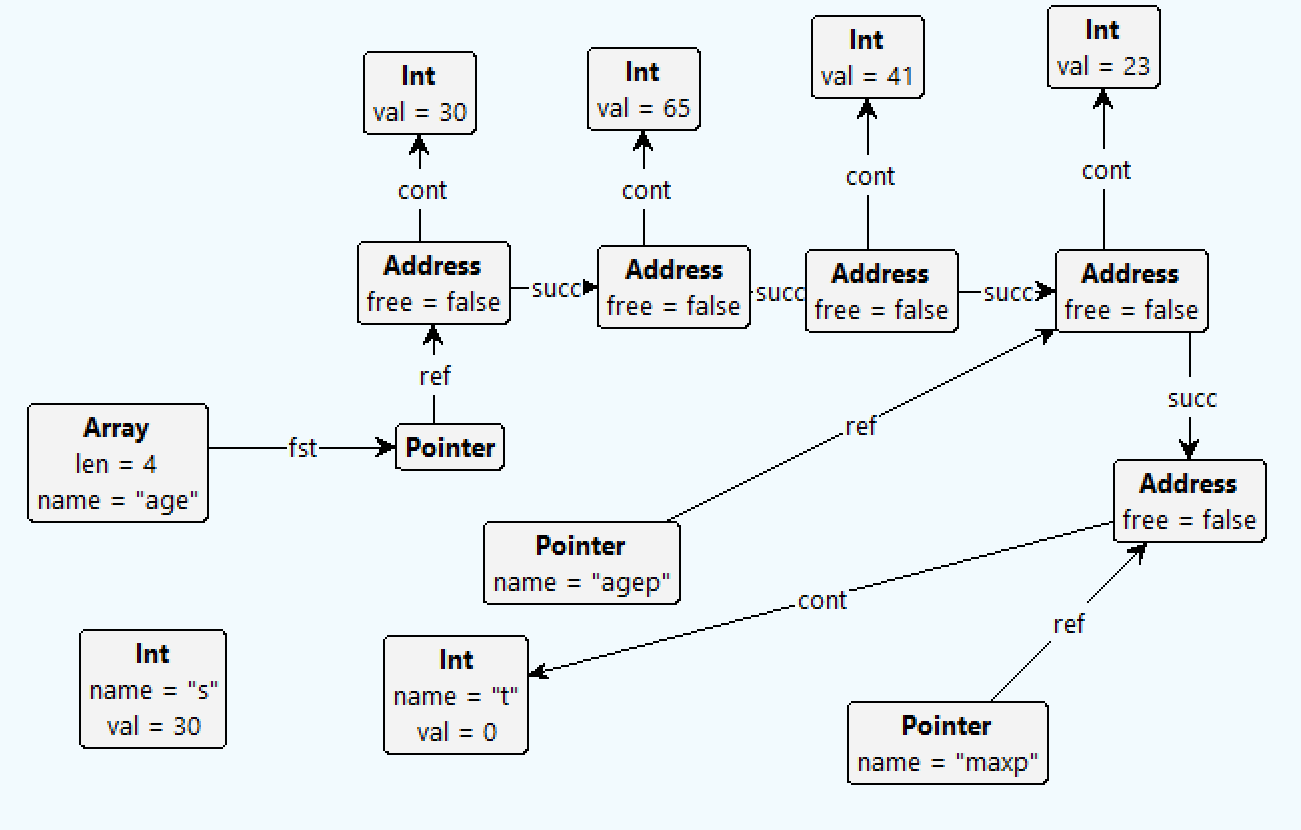}
    \end{center}
    \vspace{-20pt}
  \caption{Simulation: Fig.~\ref{nullpointerInt} applied for \texttt{*maxp=t;}}
  \label{step4-applied}
\end{figure}

A sample simulation using some of the GT rules defined in Section~\ref{pointer} is shown in Figures~\ref{step1-match}-~\ref{step4-applied}.


\section{Evaluation}\label{Experiment}
The pedagogical impact of using graphical simulations for teaching C pointer manipulation was tested on a class of students on the Introduction to C Programming course in Spring 2023. As a benchmark, we used another class of students on the same course during Spring 2022. Both sets of students were taught pointers  from the same textbook~\cite{afischer} and given the same review questions. 

The visualisations that the Spring 2023 class received included PowerPoint slides summarising the lecture on pointers, an in-class simulation demo, practice exercise and a Zoom recording of the lecture. The PowerPoint slides introduced the students to the type graph, start graph, reviewed basic pointer notation, and illustrated how the pointer notation is represented in the graph, as shown in Figure~\ref{notation}. After reviewing basic concepts, the Groove Simulator tool was used to further explore and understand the start graph, which matched a homework exercise problem (Table 1). A simulation run was then performed by illustrating different possible operations specified by rules. Once this was complete, we went back to the original homework question by adding the correct notation to the code on the whiteboard.
Both Spring classes were assessed using the same questions on their test and final exam, similar to the following textbook question~\cite{afischer}.

\smallskip
\noindent\textbf{Question:} Complete each of the following C statements by adding an \emph{asterisk \texttt{*}}, \emph{ampersand  \texttt{\&}}, or \emph{subscript \texttt{[ ]}} wherever needed to make the statement do the job described by the comment. 
Exercise question and answers are shown in Table 1. These question require a clear understanding of pointers and the notations used to operate on them.
Use these declarations:\newline
\texttt{int s=0;} \newline
\texttt{int t=0;} \newline
\texttt{int age[] = { 30, 65, 41, 23 };}\newline
\texttt{int * agep, * maxp;} 

\smallskip

\begin{table}[h!]
\begin{tabular}{|l|l|}
  \hline
   Question: Add an \texttt{*},\texttt{\&}, or \emph{subscript \texttt{[ ]}} wherever needed & Answer \\
    \hline
  \begin{minipage}{9cm}
    \vskip 4pt
    \begin{enumerate}[a.]
   \item // Make agep refer to first age in array \newline
    \texttt{agep = age;}
\item  // Copy value of agep's referent into s.  \newline
 \texttt{s = agep;}
\item  // Copy 65 into agep's referent.  \newline
\texttt{agep = age[];}
\item  // Make maxp refer to agep's referent. \newline
\texttt{maxp = agep;}
\item // Store mean of 2nd and last ages in agep's referent.  \newline
\texttt{agep = (age[]+age[])/2;} 
\item // Read into third array slot. \newline
\texttt{scanf( "\%hi", age[] ); }
\item // Read into agep's referent.  \newline
\texttt{scanf( "\%hi", agep ); }
\item // Print agep's referent.  \newline
\texttt{printf( "\%hi", agep );} 
   \end{enumerate}
   \vskip 4pt
 \end{minipage} &
 \begin{minipage}{6cm}
    \vskip 4pt
    \begin{enumerate}[a.]
    \item \texttt{agep = age;} \newline
\item \texttt{s = *agep;} \newline
\item \texttt{*agep = age[1];} \newline
\item \texttt{maxp = agep;} \newline
\item \texttt{*agep = (age[1]+age[3])/2;} \newline
\item \texttt{scanf( "\%i", \&age[2] );} \newline
\item \texttt{scanf( "\%i", agep );} \newline
\item \texttt{printf( "\%i", *agep );} 
    \end{enumerate}
   \vskip 4pt
 \end{minipage}
 \\
  \hline
 \end{tabular}
      \label{qa}
   \caption{Exercise: Question \& Answer} 
\end{table}






\subsection{Results}
We compare the test results of the three classes ( Spring 2022, 2023 \& Fall 2022) and present some observations gathered from a survey of the Spring 2023 class. A value-added score~\cite{value-added} was used to compare the test results of the three classes. 

\begin{figure}[h!]
  \begin{center}
  \includegraphics[width=13cm]{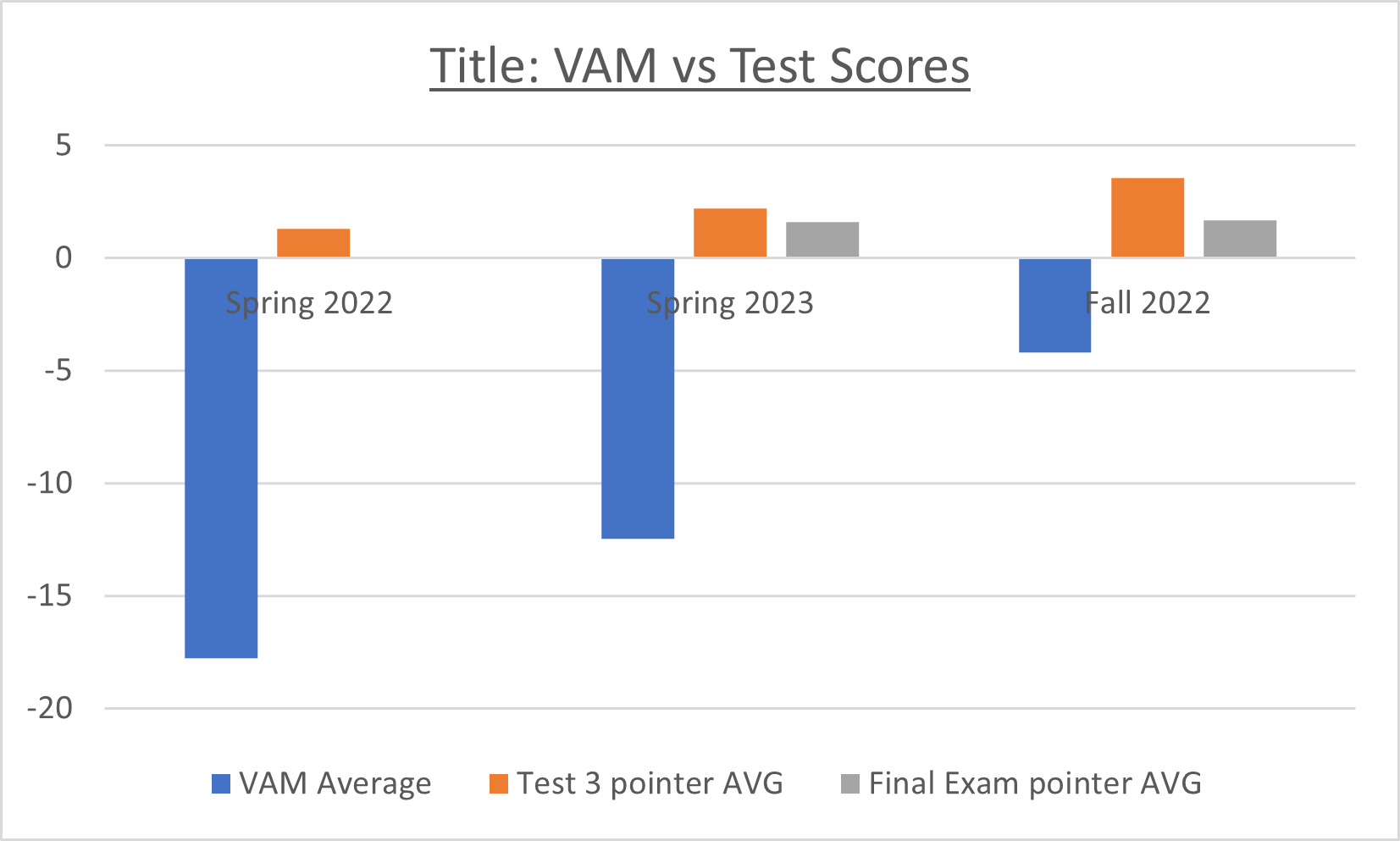}
    \end{center}
    \vspace{-20pt}
  \caption{Evaluation Result: Value Added Measure vs Pointers test results in 3 semesters }
  \label{evalresults}
\end{figure}

A detailed evaluation was done comparing Spring 2023, Spring 2022, Fall 2022 overall value added scores and pointer test scores. In each semester, each student's value added score was calculated based the expected test score from their previous two test in comparison to their actual test 3 mark. The individual students' value added scores were than averaged in each semester.  Results are as follows:
\begin{itemize}
\item Spring 2022: -17.76833333
\item Fall 2022: -4.1818
\item Spring 2023: -12.4559375
\end{itemize}
The value added measures (VAM)~\cite{vamsite} was determined based on the performance of the students in the various tests. First, the expected Test 3 scores were calculated for each individual student based on their Test 1 and 2 grades. Then the expected Test 3 score was compared to the actual score which then formed the basis of the VAM.  In all cases, the VAM scores were negative because Test 3 was harder material then the prior two tests, which therefore produced overly optimistic predictions. 
Figure ~\ref{evalresults} shows the value added measures, the pointer test scores for Test 3 and the final exam. The results show that amounts of the negative value added measures are decreasing from Spring 2022 to 2023; hence the effectiveness of the pedagogy is demonstrated. The validity of this statistical analysis is limited by the fact that students in different cohorts had different aptitudes. Spring semester student tend to be either retaking the C programming course, or they take the course later because they initially entered the university with poor math scores. However, the value-added scores shows that the pedagogy introduced in Spring 2023 positively impacted test and final exam pointer scores. 

\begin{figure}[h!]
  \begin{center}
  \includegraphics[width=10cm]{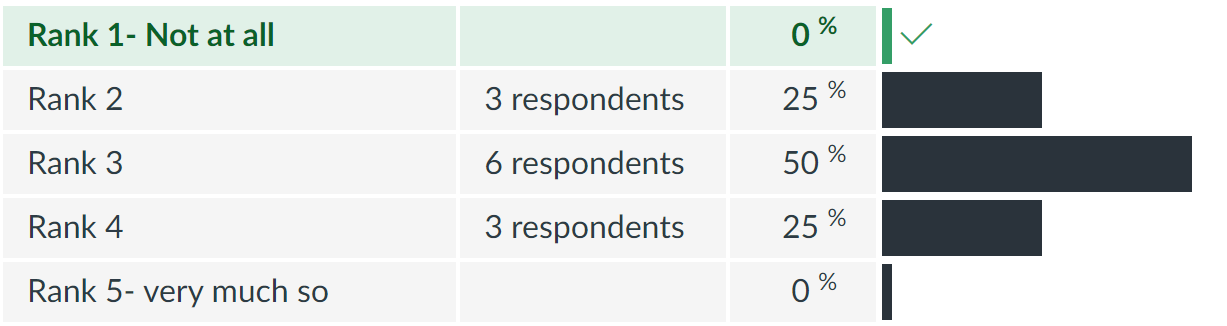}
    \end{center}
    \vspace{-20pt}
  \caption{Survey Results: Helpfulness on Pedagogy}
  \label{helpful}
\end{figure}

Figure~\ref{helpful} shows the survey results for the question: "The visual representation helped me to understand what happens when the code is executed".
Students were then asked to comment on the open-ended question: ``How did it help, what was the additional insight gained from the graphical view?''
Some of the anonymous responses were as follows:
\begin{itemize}
\item "Although it is still a slightly confusing topic for me, I realize that when trying to understand pointers it is very important to understand where everything is being stored and where things are being directed. The graphical view helped me visualize this easier. "
\item "It helped me better understand that pointers are useful in programs and the graphical view helped me visualize the flow of pointers and how they operate in a program."
\item "It was difficult to visualize how pointers worked and having a graphical view helped me understand how it pointed to different memory locations."
\item "I'm more of a visual learner, so it helped me understand it in more than one way."
\item "It helped me understand how connected it could be with the tree with all the connecting parts"
\end{itemize}

\section{Related Work}\label{Related}

While there are other novel approaches to teaching C pointers, we believe that our pedagogy provides a distinctive angle.
\cite{vocational} presents a traditional approach to teaching pointer manipulation used at a vocational college in China, which they believe to be an effective method.  The approach  distinguishes the key concepts of a memory address as name, address and content. They also used pictorial representations, and small coding examples. In addition to the traditional approach they also presented pointers in terms of analogies. For instance, a teacher office building contains room with room numbers (addresses) containing teachers (content). Otherwise~\cite{vocational} it is  similar to the traditional method used at UNH as discussed in Section~\ref{Background}. However, authors agree that many students have difficulties in understanding pointers using the classical learning approach~\cite{seriousgame} and that there is a need for novel solutions.

\cite{vtp} took on a Value Trace Problem (VTP) approach on pointers, where they provide source code, a set of questions and hints to a group of students. The questions ask the students to specify the value of a variable or output message from the code. The results are validated using string matching. Their approach was very similar to the traditional teaching with added  hints and study questions, not visual but focused on actual memory address values. For example, given code \texttt{char test1[] = {'o','k','a','y','a','m','a'};} and hint: \emph{The address of \texttt{test1[0]} is 0028FF30}, what is the address of \texttt{'test1[1]'}~\cite{vtp}. This level of detail is avoided in our approach, where addresses are abstract nodes whose numerical representation is hidden. This comes at the cost of \emph{succ} edges to represent their linear order, but supports the same concepts in relation to pointers, addresses and their operations \texttt{* ,\&}. 

\cite{industralExample} focuses on approaches to teach C pointers to mechanical engineering students, using a hydraulic press as an  example. They addressed multi-dimensional arrays with pointers to read/write values from the memory of embedded devices. The main similarity here is that the  method is also visual; however the visualisation was a depiction of a physical object. The key difference is that their their pedagogy is domain-specific, intended for mechanical engineering students while we make no such restriction.  

Another approach to teaching programming uses serious games. The Perobo serious game is inspired by the operation of computer memory~\cite{seriousgame} using a constructivist approach to encourage the learner to complete activities such as completing definitions (on variables), guessing types for pointer declarations, and matching Random Access Memory (RAM). Our approach is implementation-oriented and has a simulation component. Both approaches are visual, however in different ways and different levels of abstraction. Our visualisation is at a lower level of abstraction, focusing on object, address, and pointer relations shown as graphs. Instead, \cite{seriousgame} presents a visual game based on real-world metaphors. For instance, one of the game activities has a matrix depicting different memory locations, containing different types of robots. One of the robots represents a pointer while the others represent variables. The user is required to associate the pointer with the right address based on the values of the variables. 

Graphs and graph rewriting have also been used for the \emph{verification} of pointer programs. \emph{Shape analysis} aims to check correctness with respect to constraints demanding, for example, that a data structure has a tree or list shape. 
Representations are often at a higher level of abstraction, where objects are nodes and pointers are edges, rather than the implementation-oriented model in our approach. For example, \cite{shapetypes} propose a notion of shapes defined by a context-free graph grammar and an algorithm for static shape analysis of graph transformers. They also introduce a notation for shape types and transformers in C.  \cite{graphReduction} and \cite{shapeSafety} implement shape types by graph reduction to define and validate structures such as cyclic lists, linked lists, binary search trees, red-black trees and balanced binary tree. They have an algorithm for shape safety of operations such as search, insertion and deletion. \cite{verifyingPointer} use graph grammars to obtain finite abstractions of  pointer-manipulation programs to find bugs due to dereferencing of null pointers and memory leaks. Neither of these approaches aims at the pedagogy of teaching pointer manipulations. 


In conclusion, our approach is more abstract than using numerical pointer values while being generic rather than depicting real-world objects and scenarios, using simulations to animate graphs representing pointer structures. 
Other approaches that also represent pointer manipulation by graphs transformation have different motivations and are presented at different level of abstractions.

\section{Conclusion and Future Work}  \label{Conclusion}

We presented an initial experiment in graph transformation for teaching pointer manipulation in C and reported on our experience with a class in the Spring term of 2023. The results are promising but not conclusive because, although we accounted for prior attainment using a value added score, a range of other factors could have affected students' performance across two years. 


Our approach was re-evaluated using a more general model during the Fall 2023 semester in two C programming classes of 25 students each. Both classes were taught pointers using our new pedagogy and we compared the results to the Fall 2022 classes, which were not taught using the approach. The Fall 2023 students also had hands-on exposure to the Groove tool as independent extra credit project. We used similar pointer questions in their test and final exam to evaluate them, but due to a different model the results are incomparable to the ones presented here.

In particular, we updated the type graph in order to represent the linear address space by numerical address attributes rather than successor edges. This makes for a simpler representation, in line with how addresses work at machine level. A visualisation of pointers as edges can be derived from the numerical representation. 

The Fall 2023 course began at the end of August 2023, however the syllabus was adjusted to ensure there will be adequate time to review pointers using the updated approach and introduce the students to graph transformation concepts and the Groove tool. Hence, Chapter 1 and  Chapter 2: Programs and Programming  of~\cite{afischer} were completed in the first week of the class (rather than starting in the second week as in Spring 2023); therefore by, November the students should have completed the majority of the required chapters outlined in Section~\ref{Background}. 

In the Spring 2023 evaluation discussed in Section~\ref{Experiment}, the approach was only used  to revise pointer, and was demonstrated to the students. This Fall the approach is used in the original delivery of the material. Also, students were be provided with a small tutorial on graph transformation and Groove in order to prepare them to have interactive experiments using the tool and the updated C pointer graph transformation system. A full report on this experiment is beyond the scope of this paper because it would require the introduction of a new model and different experiemntal setup.

Further developments will include:
\begin{itemize}

\item Defining a complete set of rules for pointer manipulation in C, including memory management using \texttt{malloc()}, \texttt{calloc()}, \texttt{realloc()} and \texttt{free()} and validating these using a range of sample programs.

\item Considering function calls, including the difference between call by reference and call by value, and distinguishing stack from heap memory.

\item 
Adding referential integrity constraints to the curriculum to help explain what are desirable (consistent) pointer structures, possible inconsistencies, and how they may affect program behaviour.

\item 
Adding subject-specific visualisations for arrays and consecutive addresses to align notation to what may be found in textbooks.
\end{itemize}

\paragraph{Acknowledgment:}
We would like to thank the students of the Introduction to C Programming class at the University of New Haven, CT, USA for their participation.
\bibliographystyle{eptcs}
\bibliography{generic}
   


\end{document}